\documentclass[showpacs,preprint,amsmath,amssymb]{revtex4}
\usepackage{graphicx}
\usepackage{dcolumn}
\usepackage{epsfig}
\usepackage{bm}

\begin{document}

\title{ Neutrino-induced pion production from nuclei at medium energies }
\author{ C.~Praet }
\email{ christophe.praet@ugent.be }
\author{ O.~Lalakulich }
\altaffiliation{ Present address: Institut f\"{u}r Theoretische Physik, Universit\"{a}t
  Giessen, Germany.} 
\author{ N.~Jachowicz }
\author{ J.~Ryckebusch }
\affiliation{ Department of Subatomic and Radiation Physics,\\ Ghent
  University, \\Proeftuinstraat 86, \\ B-9000 Gent, Belgium. }

\date{\today}

 

\begin{abstract}

We present a fully relativistic formalism for describing neutrino-induced
$\Delta$-mediated single-pion production from nuclei.  We assess the
ambiguities stemming from the $\Delta$ interactions.  Variations in
the cross sections of over $10\%$ are observed, depending on whether
or not magnetic-dipole dominance is assumed to extract the vector form
factors.  These uncertainties have a direct impact on the accuracy
with which the axial-vector form factors can be extracted.  Different
predictions for $C_5^A(Q^2)$ induce up to $40$-$50\%$ effects on the
$\Delta$-production cross sections.  To describe the nucleus, we turn
to a relativistic plane-wave impulse approximation (RPWIA) using
realistic bound-state wave functions derived in the Hartree
approximation to the $\sigma$-$\omega$ Walecka model.  For neutrino
energies larger than $1$\ GeV, we show that a relativistic Fermi-gas
model with appropriate binding-energy correction produces comparable
results as the RPWIA which naturally includes Fermi motion,
nuclear-binding effects and the Pauli exclusion principle.  Including
$\Delta$ medium modifications yields a $20$ to $25\%$ reduction of the
RPWIA cross section.  The model presented in this work can be
naturally extended to include the effect of final-state interactions
in a relativistic and quantum-mechanical way.  Guided by recent
neutrino-oscillation experiments, such as MiniBooNE and K2K, and
future efforts like MINER$\nu$A, we present $Q^2$, $W$, and various
semi-inclusive distributions, both for a free nucleon and carbon,
oxygen and iron targets.               

\end{abstract}

\maketitle

\section{ Introduction }
\label{intro}

In the last few years, precision measurements of the
neutrino-oscillation parameters have driven the interest in
medium-energy neutrino physics.  The MiniBooNE \cite{MiniBooNE} and
K2K \cite{K2K} collaborations have recently collected a wealth of
neutrino data in the $1$-GeV energy range \cite{NuInt1}, where
the vast part of the strength can be attributed to quasi-elastic (QE)
processes and $\Delta$-mediated one-pion production.  A thorough
understanding of these cross sections is essential to reduce the
systematic uncertainties.  In turn, the high-statistics data from
these and future neutrino experiments like MINER$\nu$A \cite{Minerva1}
and SciBooNE \cite{SciBooNE} offer the opportunity to address a
variety of topics related to hadronic and nuclear weak physics.\\  
Various theoretical models have been developed to study one-pion
production on a free nucleon \cite{Rein, Graczyk, Olga1, Olga2,
  Hernandez}.  These efforts chiefly focus on studying 
the vector and axial-vector form factors that are introduced to
parameterize the incomplete knowledge of the $\Delta$-production
vertex.  Whereas the vector form factors can be reasonably well
determined from electroproduction data \cite{Graczyk, Olga2}, the
axial-vector ones remain troublesome due to the large error flags
present in early bubble-chamber neutrino data and sizeable model
dependencies in their analyses \cite{Hernandez, Kuzmin}.  Besides,
different theoretical calculations of the most important axial form
factor, $C_5^A(Q^2)$, reveal highly divergent pictures \cite{Barq, Alex1,
  Alex2}.  Consequently, the $Q^2$ evolution of the axial form factors and the
axial one-pion mass $M_A$ are rather poorly known.  Concerning the
$\Delta$-decay vertex, it has been established that the
traditionally-used decay couplings are not fully consistent with the 
Rarita-Schwinger field-theoretic description of the $\Delta$ particle
\cite{Pascalutsa}.  Instead, a consistent interaction can be constructed, 
which couples solely to the physical spin-3/2 part of the $\Delta$
propagator \cite{Pascalutsa}.  Since planned neutrino-scattering
experiments aim at putting further constraints on $M_A$ and the axial
form factors, it is important to assess the ambiguities related to the 
incomplete knowledge of the $\Delta$ interactions.\\
Modeling is made even more challenging by the fact that nuclei are
employed as detectors.  Thus, various nuclear effects need to be
addressed in order to make realistic cross-section predictions.
Traditionally, the Fermi motion of the nucleons inside the nucleus is
described within the relativistic Fermi gas (RFG) \cite{Horowitz,
  Alberico}.  Owing to its relative simplicity, the RFG model has been 
the preferred nuclear model in neutrino-event generators.  Going
beyond the RFG, realistic bound-state wave functions can be calculated 
within a relativistic shell model \cite{Alberico, Martinez}, or by
adopting spectral-function approaches that extend beyond the
mean-field picture \cite{Benhar1, Benhar2, Benhar3, Ankowski}.  A
comparison of these models provides insight into the nuclear-model
dependence of the computed cross sections.  Another nuclear effect
stems from the fact that the $\Delta$ properties are modified in a
medium \cite{Oset}, generally resulting in a shift of the peak
position and a collisional broadening of the width.  Finally, one must
consider the final-state interactions (FSI) of the outgoing pion and
nucleon.  To study the effect of FSI, recent efforts have resorted to
a combination of semi-classical and Monte-Carlo techniques
\cite{Leitner1, Singh}.  Based on these results, it is clear that FSI
mechanisms produce by far the largest nuclear effect on one-pion
production computations.\\  
In this work, we present a fully relativistic formalism that can serve 
as a starting point to investigate $\Delta$-mediated one-pion production
from nuclei.  Recognizing the ability of the new generation of
experiments to measure both inclusive and semi-inclusive observables,
we develop a framework that is geared towards a detailed study of
various distributions, like $Q^2$, $W$, energy and scattering-angle
dependences.  To model nuclear effects, we turn to the relativistic
plane-wave impulse approximation, using relativistic bound-state wave
functions that are calculated in the Hartree approximation to the
$\sigma$-$\omega$ Walecka model \cite{Furnstahl}.  This approach was
successfully applied in QE nucleon-knockout studies \cite{Martinez,
  Lava, Praet, Jachowicz}, and includes the
effects of Fermi motion, nuclear binding and the Pauli exclusion
principle in a natural way.  Medium modifications of
the $\Delta$ particle are taken into account along the lines of
Ref.~\cite{Oset}.  We investigate the sensitivity of the cross section
to uncertainties in the $\Delta$ couplings.  Then, we proceed with the
nuclear-model dependence of our results, by comparing with RFG
calculations.  Our findings in this regard are of great importance to
neutrino experiments that employ the RFG model in the event
generators.  The formalism outlined in this work is an ideal starting
ground to implement also FSI effects in a fully relativistic and 
quantum-mechanical way.  The discussion of FSI mechanisms, however,
falls beyond the scope of the present paper.\\                 
The paper is organized as follows.  Section \ref{free} introduces the
formalism for the elementary $\Delta$-mediated one-pion production
process.  The third section deals with the nuclear model and discusses
the framework for the description of neutrino-nucleus interactions.
Numerical results are presented in section IV.  In section V, we
summarize our conclusions.
                  
\section{ Charged-current pion neutrinoproduction on the nucleon }
\label{free}

\subsection{ Cross section } 
\label{cs-free}

\begin{figure}[t]
\includegraphics[width=9cm]{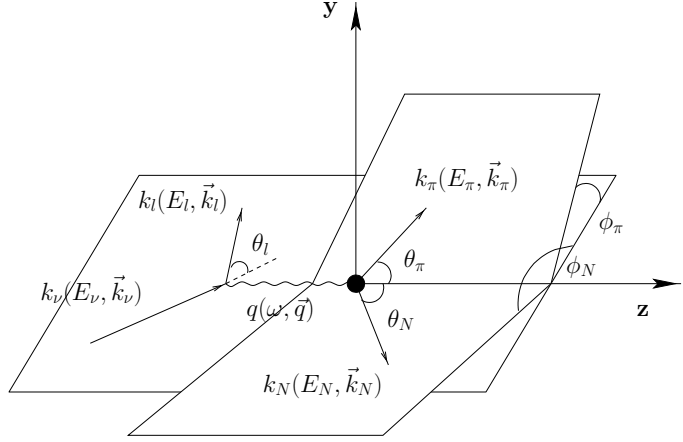}
\caption{ Kinematics for neutrino-induced charged-current one-pion
  production on the nucleon. }
\label{kinematics}
\end{figure}

For a free proton target, the charged-current (CC) process under
study is
\begin{equation}
\label{process1}
\nu_{\mu} + p \stackrel{ \Delta^{++} }{ \rightarrow } \mu^{-} + p +
\pi^{+}.
\end{equation}
The corresponding reactions for a free neutron are
\begin{equation}
\label{process2}
\begin{split}
\nu_{\mu} + & n \stackrel{ \Delta^{+} }{ \rightarrow } \mu^{-} + p +
\pi^{0},\\
\nu_{\mu} + & n \stackrel{ \Delta^{+} }{ \rightarrow } \mu^{-} + n +
\pi^{+}.
\end{split}
\end{equation}
Isospin considerations allow one to relate the strength of the above
reactions   
\begin{equation}
\label{extra1}
\begin{split}
\sigma(W^{+} p \stackrel{ \Delta^{++} }{ \rightarrow } p \pi^{+}) & =
9 \sigma(W^{+} n \stackrel{ \Delta^{+} }{ \rightarrow } n \pi^{+})\\
 & = \frac{9}{2} \sigma(W^{+} n \stackrel{ \Delta^{+} }{ \rightarrow } p \pi^{0}),
\end{split}
\end{equation}
where $W^{+}$ denotes the exchanged weak vector boson.  In a
laboratory frame of reference, the corresponding differential cross
section is given by \cite{BjDr} 
\begin{equation}
\label{free1}
\begin{split}
d^9\sigma = & \frac{ 1 }{ \beta }\frac{ m_{\nu} }{ E_{\nu} } \frac{ m_{l} }{ E_{l} }\frac{ d^3\vec{k}_{l} }{ (2\pi)^3 } \frac{
  m_{N} }{ E_{N} }\frac{ d^3\vec{k}_{N} }{(2\pi)^3} \frac{
  d^3\vec{k}_{\pi} }{ 2E_{\pi} (2\pi)^3 }\\
 \times \overline{\sum}_{fi} & |M^{(free)}_{fi}|^2 (2\pi)^4 \delta^{(4)} (k_{\nu} + k_{N,i} - k_{l} -
  k_{\pi} - k_{N}).
\end{split}
\end{equation}
Figure \ref{kinematics} defines our conventions for the kinematical
variables.
The target nucleon has four-momentum $k_{N,i} = (m_N,\vec{0})$,
with $m_N$ the nucleon's mass.  We write $k_{\nu} =
(E_{\nu},\vec{k}_{\nu})$ for the incoming neutrino, $k_{l} =
(E_{l},\vec{k}_{l})$ for the outgoing muon, $k_{\pi} =
(E_{\pi},\vec{k}_{\pi})$ for the outgoing pion and $k_{N} =
(E_{N},\vec{k}_{N})$ for the outgoing nucleon.  The $xyz$ coordinate
system is chosen such that the $z$ axis lies along the momentum
transfer $\vec{q}$, the $y$ axis along $\vec{k}_{\nu} \times
\vec{k}_{l}$, and the $x$ axis in the lepton-scattering plane.  In
Eq.~(\ref{free1}), the incoming neutrino's relative velocity $\beta =
|\vec{k}_{\nu}|/E_{\nu}$ is $1$.  The neutrino mass $m_{\nu}$ will
cancel with the neutrino normalization factor appearing in the lepton tensor.  
The $\delta$-function expresses energy-momentum conservation 
and $\overline{\sum}_{fi} |M^{(free)}_{fi}|^2$ denotes the squared
invariant matrix element, appropriately averaged over initial spins
and summed over final spins.  Using the $\delta$-function to integrate
over the outgoing nucleon's three-momentum and the magnitude of the
pion's momentum, one arrives at the fivefold cross section
\begin{equation}
\label{free3}
\begin{split}
\frac{ d^5 \sigma }{ dE_l d\Omega_l d\Omega_{\pi}} = & \frac{ m_{\nu}m_l |\vec{k}_l|
  m_{N} |\vec{k}_{\pi}| }{ 2 (2\pi)^5 E_{\nu}| E_{N} + E_{\pi}(
  |\vec{k}_{\pi}|^2 - \vec{q} \cdot \vec{k}_{\pi})/|\vec{k}_{\pi}|^2
  |} \\
 & \times \overline{\sum}_{fi} |M^{(free)}_{fi}|^2,
\end{split}
\end{equation}
where the solid angles $\Omega_l$ and $\Omega_{\pi}$ define the direction of
the outgoing muon and pion respectively.

\subsection{ Matrix element for resonant one-pion production }
\label{me-free}

\begin{figure}[ht]
\includegraphics[width=9cm]{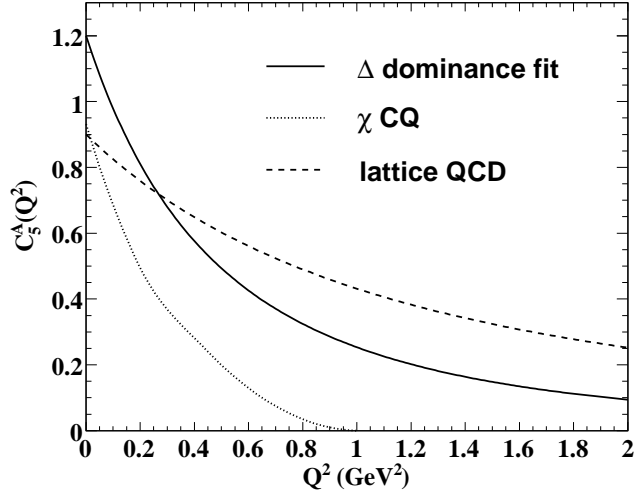}
\caption{ Different results for the axial transition form factor
  $C_5^A(Q^2)$.  The full line represents a phenomenological fit to
  ANL and BNL data within a $\Delta$-dominance model
  (Eq.~\ref{free13}).  The dashed line shows a quenched lattice
  result, and is parameterized as
  $C_5^A(Q^2)=C_5^A(0)(1+Q^2/\tilde{M}_A^2)^{-2}$, $C_5^A(0) = 0.9$\ and $\tilde{M}_A =
  1.5$\ GeV \cite{Alex2}.  The dotted line corresponds to a $\chi$CQ
  result, and is taken from Ref.~\cite{Barq}. }
\label{axial}
\end{figure}

Next to the kinematic phase-space factor, Eq.~(\ref{free3}) contains
the squared invariant matrix element 
\begin{equation}
\label{free4}
\overline{\sum}_{fi} |M^{(free)}_{fi}|^2 = \frac{1}{2} \sum_{\substack{s_{\nu};
    s_l\\ s_{N,i}; s_N}} \left[M^{(free)}_{fi}\right]^{\dagger} M^{(free)}_{fi}.
\end{equation}  
Here, the sum over final muon and nucleon spins is taken.  
Averaging over the initial nucleon's spin, $s_{N,i}$\,, leads to a
factor $1/2$.  An explicit expression for the invariant matrix element
is obtained by applying the Feynman rules in momentum space.  Writing 
\begin{equation}
\label{free5}
M^{(free)}_{fi} = i \frac{G_F \cos \theta_c}{\sqrt{2}} <J^{\rho(free)}_{had}>
S_{W,\rho\sigma} <J^{\sigma}_{lep}>,
\end{equation}
with $G_F$ the Fermi constant and $\theta_c$ the Cabibbo angle, one
distinguishes the hadron current 
\begin{equation}
\label{free6}
<J^{\rho(free)}_{had}> = \overline{u}(k_N, s_N) \Gamma^{\mu}_{\Delta
  \pi N} S_{\Delta,\mu\nu} \Gamma^{\nu\rho}_{WN\Delta} u(k_{N,i}, s_{N,i}), 
\end{equation}   
the weak boson propagator
\begin{equation}
\label{free7}
S_{W,\rho\sigma} = \frac{g_{\rho\sigma} M^2_W}{Q^2 + M^2_W};\quad
Q^2 = - q_{\mu} q^{\mu},
\end{equation}
and the lepton current
\begin{equation}
\label{free8}
<J^{\sigma}_{lep}> = \overline{u}(k_l, s_l) \gamma^{\sigma} (1 - \gamma_5)
u(k_{\nu}, s_{\nu}).
\end{equation}
Clearly, the least-known physics is contained in the
vertex functions of the matrix element of Eq.~(\ref{free6}).  For the
$\Delta$-production vertex, we adopt the form \cite{Olga1}
\begin{widetext}
\begin{equation}
\label{free9}
\begin{split}
\Gamma^{\nu\rho}_{WN\Delta}(k_{\Delta},q) & = \left[
  \frac{C_3^V(Q^2)}{m_N}(g^{\nu\rho}\not q - q^{\nu} \gamma^{\rho}) +
  \frac{C_4^V(Q^2)}{m_N^2}(g^{\nu\rho}q \cdot k_{\Delta} - q^{\nu}k_{\Delta}^{\rho}) + \frac{C_5^V(Q^2)}{m_N^2}(g^{\nu\rho}q \cdot k_{N,i} - q^{\nu}k^{\rho}_{N,i}) +
  g^{\nu\rho}C_6^V(Q^2) \right] \gamma_5\\
 & + \frac{C_3^A(Q^2)}{m_N}(g^{\nu\rho}\not q - q^{\nu}\gamma^{\rho}) +
  \frac{C_4^A(Q^2)}{m_N^2}(g^{\nu\rho}q \cdot k_{\Delta} - q^{\nu} k_{\Delta}^{\rho}) + C_5^A(Q^2) g^{\nu\rho} + \frac{C_6^A(Q^2)}{m_N^2} q^{\nu} q^{\rho},
\end{split}
\end{equation}
\end{widetext}
where a set of vector ($C_i^V,\ i=3..6$) and axial ($C_i^A,\ i=3..6$)
form factors are introduced.  These form factors are constrained by
physical principles and experimental data.  Imposing
the conserved vector current (CVC) hypothesis leads to $C_6^V = 0$.
The partially-conserved axial current (PCAC) hypothesis, together with
the pion-pole dominance assumption, yields the following relation
between $C_5^A$ and the pseudoscalar form factor $C_6^A$
\begin{equation}
\label{free10}
C_6^A = C_5^A \frac{ m_N^2 }{ Q^2 + m^2_{\pi} }.
\end{equation} 
At $Q^2 = 0$, the off-diagonal Goldberger-Treiman relation gives
$C_5^A = 1.2$ \cite{Olga2}.  Furthermore, CVC entails that the weak vector current and the
isovector part of the electromagnetic current are components of the
same isospin current.  Consequently, after extracting the
electromagnetic form factors from electroproduction data, the $C_i^V,
i=3,4,5$ follow immediately by applying the appropriate
transformations in isospin space.  To extract the vector form factors,
it has been established that the magnetic-dipole (M1) dominance of
the electromagnetic $N\rightarrow \Delta$ transition amplitude is a
reasonable assumption \cite{Stoler1}.  This M1 dominance leads to the
conditions \cite{Paschos1}
\begin{equation}
\label{free11-1}
C_4^V = - C_3^V \frac{ m_N }{ W },\quad C_5^V = 0,
\end{equation}
where $W$ is the invariant mass, defined as $W=\sqrt{k_{\Delta}^2}$.
For $C_3^V$, a modified-dipole parameterization is extracted \cite{Paschos1, Olga1, Leitner1} 
\begin{equation}
\label{free11-2}
C_3^V = \frac{ 1.95 D_V}{ 1+Q^2/4 M_V^2},
\end{equation}    
with $D_V = (1+Q^2/M_V^2)^{-2}$ the dipole function and $M_V =
0.84\ \mbox{GeV}$.  In Eq.~(\ref{free11-2}), the faster-than-dipole
fall-off reflects the fact that the $\Delta$ is a more extended object 
than a nucleon.  Within this scheme, it is possible to relate
all weak vector form factors to $C_3^V$.  More recently, a direct
analysis of the electroproduction helicity amplitudes from JLab and Mainz
experiments resulted in an alternative parameterization of the weak
vector form factors \cite{Olga2}
\begin{equation}
\label{free12}
\begin{split}
C_3^V & = \frac{ 2.13 D_V }{ 1+Q^2/4 M_V^2 },\quad C_4^V = \frac{
  -1.51 }{ 2.13 } C_3^V, \\
C_5^V & = \frac{ 0.48 D_V }{ 1+Q^2/0.776 M_V^2 },
\end{split}
\end{equation}
attributing a non-zero strength to the weak vector form factor
$C_5^V$.  The axial form factors are even more difficult to determine,
in the sense that they are only constrained by the bubble-chamber
neutrino data.  A widely used parameterization is given
by \cite{Paschos1, Olga1, Leitner1} 
\begin{equation}
\begin{split}
\label{free13}
C_5^A = & \frac{ 1.2 }{ (1+Q^2/M_A^2)^2 } \frac{ 1 }{ 1+Q^2/3 M_A^2
},\\
C_4^A = & - \frac{ C_5^A }{ 4 },\quad C_3^A = 0,
\end{split}
\end{equation}   
where $M_A = 1.05$\ GeV.  However, there still resides a great deal of
uncertainty in the axial form factors or, equivalently, in $C_5^A$.
The extracted axial-mass value, for example, is heavily
model-dependent \cite{Kuzmin, Hernandez}.  A re-analysis
\cite{Hernandez} of ANL data within a model that includes background
contributions next to the $\Delta$-pole mechanism reveals a $C_5^A(0)$
value that is lower than the one predicted by the Goldberger-Treiman
relation.  This result is corroborated by some recent form-factor calculations, within a chiral
constituent-quark ($\chi$CQ) model \cite{Barq} and lattice QCD
framework \cite{Alex1, Alex2}.  Figure \ref{axial} compares both theoretical calculations
with the phenomenological fit of Eq.~(\ref{free13}).  It can be
clearly seen that all three approaches exhibit highly divergent $Q^2$
evolutions.\\     
The Rarita-Schwinger spin-3/2 propagator for the $\Delta$ reads
\begin{widetext}
\begin{equation}
\label{free14}
S_{\Delta,\mu\nu}(k_{\Delta}) = \frac{\not{k_{\Delta}} +
  M_{\Delta}}{k^2_{\Delta} - M^2_{\Delta} + iM_{\Delta}\Gamma} \left( g_{\mu\nu} - \frac{\gamma_{\mu}\gamma_{\nu}}{3} - \frac{2 k_{\Delta,\mu}k_{\Delta,\nu}}{3M^2_{\Delta}} - \frac{\gamma_{\mu}
  k_{\Delta,\nu} - \gamma_{\nu} k_{\Delta,\mu}}{3 M_{\Delta}} \right),
\end{equation}
\end{widetext}
where $M_{\Delta} = 1.232\ \mbox{GeV}$ and $\Gamma$ stands for the
free decay width.\\
A common way of describing the $\Delta$ decay is through the interaction
Lagrangian
\begin{equation}
\label{free15}
\mathcal{L}_{\pi N \Delta} = \frac{ f_{\pi N \Delta} }{ m_{\pi} }
\overline{\psi}_{\mu} \vec{T}^{\dagger} (\partial^{\mu} \vec{\phi})
\psi + h.c, 
\end{equation}
where $\psi_{\mu}$, $\vec{\phi}$ and $\psi$ denote the spin-3/2
Rarita-Schwinger field, the pion field and the nucleon field
respectively.  The operator $\vec{T}$ is the isospin 1/2 $\rightarrow$
3/2 transition operator.  From (\ref{free15}), one derives the simple      
vertex function
\begin{equation}
\label{free16}
\Gamma^{\mu}_{\Delta \pi N}( k_{\pi} ) = \frac{ f_{\pi N \Delta} }{
  m_{\pi} } k^{\mu}_{\pi},
\end{equation}
and the corresponding energy-dependent width
\begin{equation}
\label{free17}
\Gamma(W) = \frac{1}{12 \pi} \frac{ f^{2}_{\pi N \Delta} }{ m^2_{\pi} W }
|\vec{q}_{cm}|^3 ( m_N + E_N ),
\end{equation}
with
\begin{equation}
| \vec{q}_{cm} | = \frac{ \sqrt{(W^2 - m^2_{\pi} - m_N^2)^2 - 4m_{\pi}^2 m_N^2} }{ 2 W }.
\end{equation}
Requiring that $\Gamma(M_{\Delta})$ equals the experimentally determined
value of $120$ MeV, one obtains $f_{\pi N \Delta} = 2.21$.  An
alternative choice for the $\Delta \pi N$ interaction Lagrangian is
provided by
\begin{equation}
\label{free18}
\mathcal{L}_{\pi N \Delta} = \frac{ f^*_{\pi N \Delta} }{ m_{\pi}
  M_{\Delta} } \epsilon^{\alpha\beta\mu\nu} \overline{G}_{\beta\alpha}
  \gamma_{\mu} \gamma_5 \vec{T}^{\dagger} (\partial_{\nu} \vec{\phi})
  \psi, 
\end{equation}   
where $G_{\beta\alpha} = \partial_{\beta}\psi_{\alpha} -
\partial_{\alpha}\psi_{\beta}$.  This form has been
proposed by Pascalutsa et al. \cite{Pascalutsa}, who point out that many of the
traditional couplings, like the one in Eq.~(\ref{free15}), give rise to
unwanted spin-1/2 contributions to the cross section.  The interaction
of Eq.~(\ref{free18}), however, couples only to the
physical, spin-3/2 part of the $\Delta$ propagator.  With the
interaction Lagrangian of Eq.~(\ref{free18}) the vertex function becomes
\begin{equation}
\label{free19}
\Gamma^{\mu}_{\Delta \pi N}(k_{\pi},k_{\Delta}) = \frac{f^*_{\pi N \Delta}}{m_{\pi}
  M_{\Delta}} \epsilon^{\mu \alpha \beta \gamma} k_{\pi, \alpha}
  \gamma_{\beta} \gamma_5 k_{\Delta,\gamma}.
\end{equation}
Calculating the decay width from Eq.~(\ref{free19}) leads to the same
expression as in Eq.~(\ref{free17}), implying $f^*_{\pi N \Delta} = f_{\pi N
  \Delta} = 2.21$.\\
Combining formulas (\ref{free4}) to (\ref{free8}), the squared
invariant matrix element can be cast in the form
\begin{equation}
\label{free21}
\overline{\sum}_{fi} |M^{(free)}_{fi}|^2 = \frac{ G_F^2 \cos^2\theta_c M_W^4 }{
2 (M_W^2 + Q^2)^2 } H^{\rho\sigma}_{(free)} L_{\rho\sigma},
\end{equation}
where the leptonic tensor is given by
\begin{equation}
\label{free22}
L_{\rho \sigma} = \frac{2}{m_{\nu} m_l} ( k_{\nu,\rho}
 k_{l,\sigma} + k_{\nu,\sigma} k_{l,\rho} - k_{\nu}\cdot k_l
 g_{\rho\sigma} - i\epsilon_{\alpha\rho\beta\sigma} k_{\nu}^{\alpha}
 k_l^{\beta} ),
\end{equation}
with the definition $\epsilon_{0123} = +1$.  Introducing the shorthand
notation $\mathcal{O}^{\sigma} = \Gamma^{\mu}_{\Delta\pi N}
S_{\Delta,\mu\nu} \Gamma^{\nu\sigma}_{WN\Delta}$, one arrives at the
following expression for the hadronic tensor
\begin{equation}
\label{free23}
 H^{\rho\sigma}_{(free)} = \frac{1}{8 m^2_N} \mbox{Tr}\Big( (\not k_{N,i} + m_N) \widetilde{\mathcal{O}}^{\rho} (\not k_N + m_N) \mathcal{O}^{\sigma} \Big),
\end{equation}
where $\widetilde{\mathcal{O}}^{\rho} = \gamma_0
(\mathcal{O}^{\rho})^{\dagger} \gamma_0$.

\section{ Charged-current pion neutrinoproduction from a nucleus }
\label{bound}

Turning to nuclear targets, a schematical representation of the
reaction under study is given by  
\begin{equation}
\label{bound-cs1}
\nu_{\mu} + A \stackrel{ \Delta }{ \rightarrow } \mu^{-} + (A-1)
+ N + \pi,
\end{equation}
where $A$ denotes the mass number of the target nucleus.  Compared to 
the free-nucleon case, one now needs to consider the residual nucleus
$k_{A-1} = (E_{A-1},\vec{k}_{A-1})$ as an extra particle
in the hadronic final state.  Following the same line of reasoning as
in section \ref{cs-free}, the lab-frame cross section corresponding to
the process of Eq.~(\ref{bound-cs1}) becomes
\begin{equation}
\label{bound-cs2}
\begin{split}
 & \frac{d^8\sigma}{dE_l d\Omega_l dE_{\pi} d\Omega_{\pi} d\Omega_{N}} = \\ 
 & \frac{m_{\nu}m_l |\vec{k}_l| m_N m_{A-1} |\vec{k}_{\pi}| |\vec{k}_N|}{2 (2\pi)^8 E_{\nu}|E_{A-1} + E_N + E_N \vec{k}_N \cdot
  (\vec{k}_{\pi} - \vec{q})/|\vec{k}_N|^2|} \\
 & \hspace{.2\textwidth} \times \overline{\sum}_{fi} |M^{(bound)}_{fi} |^2.
\end{split}
\end{equation}

\subsection{Relativistic bound-state wave functions}

The invariant matrix element in (\ref{bound-cs2}) carries the tag
\textit{bound} and involves nuclear many-body currents between
initial and final nuclear wave functions.  In medium-energy physics,
however, one usually resorts to a number of assumptions that allow a 
reduction of the nuclear-current matrix elements to a form similar to
Eq.~(\ref{free6}).  Here, we summarize the main approximations that
enable this simplification and refer to Ref.~\cite{Cosyn1} for the
more detailed and analytic considerations.  First, we only consider
processes where the residual $(A-1)$ system is left with an excitation
energy not exceeding a few tens of MeV.  The major fraction of the
transferred energy is carried by the outgoing pion and nucleon.
Further, we adopt the impulse approximation (IA): the nuclear
many-body current is replaced by a sum of one-body current operators, 
exempt from medium effects.  Assuming an
independent-particle model (IPM) for the initial and final nuclear
wave functions, the hadronic current matrix elements can be
written in the form of Eq.~(\ref{free6}), whereby the
initial-nucleon free Dirac spinor is replaced by a bound-state
spinor \cite{Cosyn1}.  This approach, where the outgoing nucleon and
pion remain unaffected by the nuclear medium, is generally referred to
as the relativistic plane-wave impulse approximation (RPWIA).\\
The single-particle wave functions used in this work are determined in
the Hartree approximation to the $\sigma$-$\omega$ Walecka model,
using the $W1$ parameterization for the different field strengths \cite{Furnstahl}.
They are written as 
\begin{equation}
\label{bswf1}
\Psi_{\alpha,m}(\vec{r}) = \left( \begin{array}{c} i
  \frac{G(r)}{r} \mathcal{Y}_{+\kappa,m}(\hat{\vec{r}})  \\ 
- \frac{F(r)}{r} \mathcal{Y}_{-\kappa,m}(\hat{\vec{r}})
\end{array} \right),
\end{equation}
where $m$ is the magnetic quantum number and $\alpha$ stands for all
other quantum numbers that specify a single-particle orbital.  In the
definition of the spherical two-spinors, a generalized angular
momentum $\kappa$ is introduced.  The momentum wave functions are
obtained from
\begin{equation}
\label{bswf2}
\mathcal{U}_{\alpha,m}(\vec{p}) = \frac{1}{(2\pi)^{3/2}} \int
\Psi_{\alpha,m}(\vec{r}) e^{-i\vec{p}\cdot \vec{r}} d\vec{r}.
\end{equation}
The result is
\begin{equation}
\label{bswf3}
\mathcal{U}_{\alpha,m}(\vec{p}) = i^{(1-l)} \sqrt{\frac{2}{\pi}} \frac{1}{p} \left( \begin{array}{c} 
  g(p) \mathcal{Y}_{+\kappa,m}(\hat{\vec{p}})  \\ 
- f(p) \mathcal{Y}_{-\kappa,m}(\hat{\vec{p}})
\end{array} \right),
\end{equation}
with 
\begin{equation}
\label{bswf4}
g(p) = \int_0^{\infty} G(r) \hat{\jmath}_l(pr) dr,
\end{equation} 
and
\begin{equation}
\label{bswf5}
f(p) = \mbox{sgn}(\kappa) \int_0^{\infty} F(r)
\hat{\jmath}_{\overline{l}}(pr) dr, \quad \overline{l} = \left( \begin{array}{c} 
  l + 1,\quad \kappa < 0  \\ 
  l - 1,\quad \kappa > 0 \end{array} \right).
\end{equation}
In (\ref{bswf4}) and (\ref{bswf5}), $\hat{\jmath}_l(x) = x \, j_l(x) $
is the Ricatti-Bessel function.\\
Returning to the calculation of the squared invariant matrix element
in (\ref{bound-cs2}), the following factor appears   
\begin{equation}
\label{bswf6}
S_{\alpha}(\vec{p}) = \frac{1}{2j+1} \sum_m \mathcal{U}_{\alpha,m}(\vec{p})
\mathcal{\overline{U}}_{\alpha,m}(\vec{p}).
\end{equation}   
This expression, referred to as the bound-state propagator, can be
cast in a form which is similar to the free-nucleon projection
operator \cite{Gardner}.  One finds
\begin{equation}
\label{bs-prop}
S_{\alpha}(\vec{p}) = ( \not k_{\alpha} + M_{\alpha} ),
\end{equation} 
with the definitions
\begin{equation}
\label{bswf7}
\begin{split}
M_{\alpha} & = \frac{1}{(2\pi)^3}\frac{\pi}{p^2} \left( g^2(p) - f^2(p) \right),\\
E_{\alpha} & = \frac{1}{(2\pi)^3}\frac{\pi}{p^2} \left( g^2(p) + f^2(p)\right),\\
\vec{k}_{\alpha} & = \frac{1}{(2\pi)^3}\frac{\pi}{p^2} \left( 2 g(p) f(p) \hat{\vec{p}} \right).
\end{split}
\end{equation}
\begin{figure}[t]
\includegraphics[width=18cm]{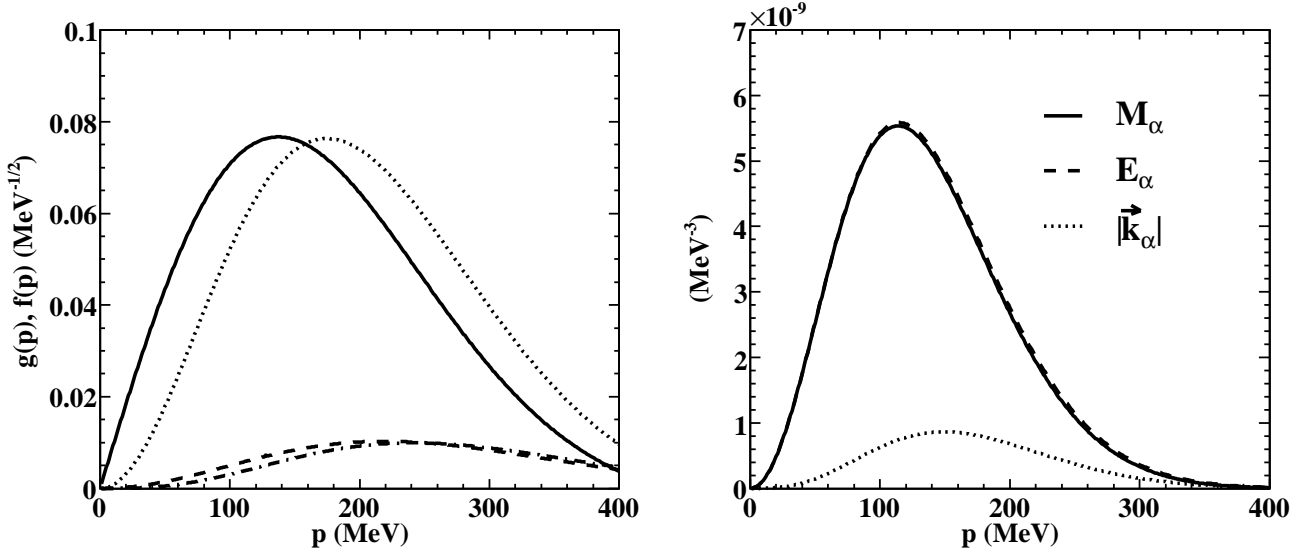}
\caption{ The left panel shows the momentum wave functions for the
  carbon nucleus.  The full (dashed) line corresponds to $g(p)$
  ($f(p)$) for a $1s_{1/2}$ proton, the dotted (dash-dotted) line represents $g(p)$
  ($f(p)$) for a $1p_{3/2}$ proton.  In the right panel, the quantities
  defined in Eq.~(\ref{bswf7}) are shown for a $1p_{3/2}$-shell
  $^{12}\mathrm{C}$ proton. }
\label{bswf-fig}
\end{figure}
In other words, the hadronic tensor for scattering off a bound nucleon
is readily found from the free-nucleon one in Eq.~(\ref{free23}) by making
the replacement 
\begin{equation}
\label{bswf8}
\frac{1}{2} \frac{(\not k_{N,i} + m_N)}{2m_N} \longrightarrow (2\pi)^3
( \not k_{\alpha} + M_{\alpha} ).
\end{equation}
Figure \ref{bswf-fig} shows the momentum wave functions of
Eqs.~(\ref{bswf4}) and (\ref{bswf5}) for a proton belonging to a specified carbon
shell.  Owing to the small contribution of the lower wave-function component,
the quantities $M_{\alpha}$ and $E_{\alpha}$ are almost equal in
strength. 

\subsection{ Medium modifications of $\Delta$ properties }
\label{MMs}

In a nuclear environment, the $\Delta$ mass and width will be
modified with respect to its free values.  These medium modifications
can be estimated by calculating the in-medium $\Delta$ self-energy, as
was e.g. done in Ref.~\cite{Oset}.  The
real part of the $\Delta$ self-energy causes a shift of the resonance
position, whereas the imaginary part is related to the decay width.  Medium modifications
for the width result from the competition between a Pauli-blocking
correction, reducing the free decay width, and a term proportional to
the imaginary part of the $\Delta$ self-energy, including various
meson and baryon interaction mechanisms and, therefore, enhancing the
free decay width.  A convenient parameterization for
the medium-modified mass and width of the $\Delta$ is given in
Ref.~\cite{Oset}, in terms of the nuclear density $\rho$.  For our
purposes, we shall adopt an average nuclear density $\rho = 0.75
\rho_0$, with $\rho_0$ the equilibrium density.  Then, at the $\Delta$
peak, we calculate the following shifts 
\begin{equation}
\label{MM1}
\begin{split}
M_{\Delta} & \longrightarrow M_{\Delta} + 30\ \mbox{MeV},\\
\Gamma & \longrightarrow \Gamma + 40\ \mbox{MeV}.
\end{split}
\end{equation}    
In Ref.~\cite{MacGregor}, a similar recipe was used to accommodate
medium modifications in the calculation of
$^{12}\mathrm{C}(\gamma,\mathrm{pn})$ and
$^{12}\mathrm{C}(\gamma,\mathrm{pp})$ cross sections.  There, the
computations proved to compare favorably with the data in an energy regime
where the reaction is dominated by $\Delta$ creation.                

\section{ Results and discussion }
\label{results}

In this section, we present computations for the process
\begin{equation}
\label{res1}
\nu_{\mu} + p \stackrel{ \Delta^{++} }{ \rightarrow } \mu^{-} + p +
\pi^{+},
\end{equation}  
the strength of which can be straightforwardly related to the other
channels listed in Eq.~(\ref{process2}) by applying the isospin
relations of Eq.~(\ref{extra1}).  Throughout this work, the cross
sections are shown per nucleon, meaning that the RPWIA results are
scaled to the number of protons in the target nucleus.  Unless
otherwise stated, we use the vector form factors derived in the
M1-dominance model (Eqs.~(\ref{free11-1}) and (\ref{free11-2})), 
the axial form factors of Eq.~(\ref{free13}) with $M_A =
1.05\ \mbox{GeV}$, and the \textit{traditional} $\Delta\pi N$
coupling defined in Eq.~(\ref{free15}).  For the RFG calculations, 
we adopt $k_F = 225\ \mbox{MeV}$ and an average binding energy of 
$E_B = 20\ \mbox{MeV}$.  The latter value can be considered as a good
estimate for the weighted average of the centroids of the
single-particle strength distributions in typical even-even nuclei
near the closed shells \cite{Heyde}. 

\subsection{WN$\Delta$ and $\Delta\pi$N couplings}
\label{results-pt1}

\begin{figure}[t]
\includegraphics[width=18cm]{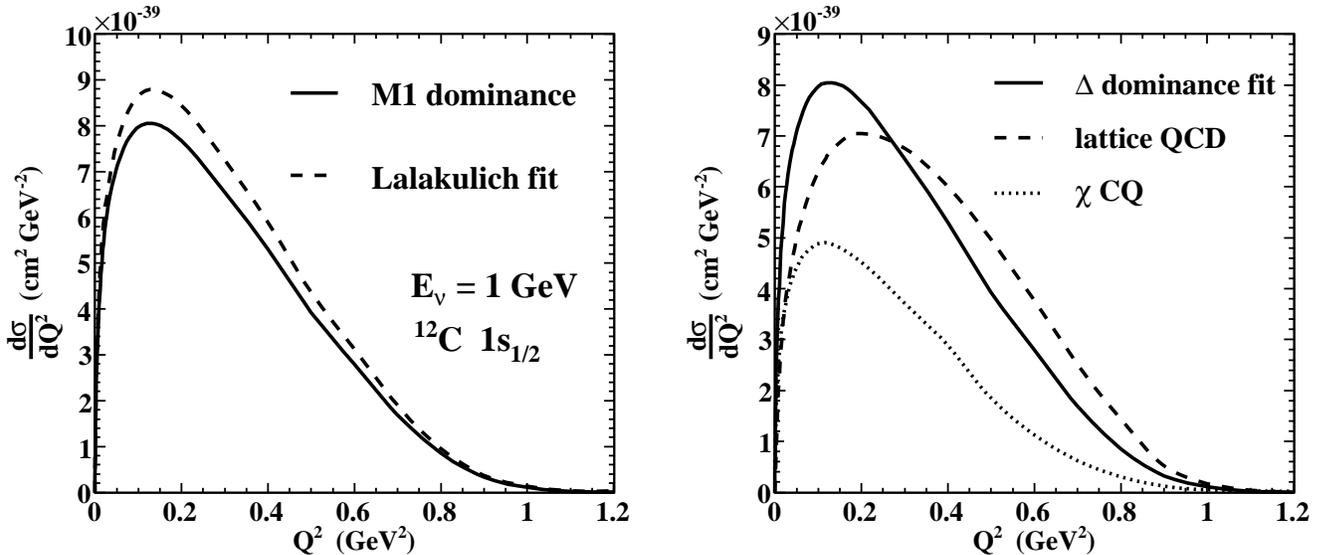}
\caption{ $Q^2$ evolution of the $\Delta^{++}$-production cross
  sections for a $1s_{1/2}\ ^{12}\mbox{C}$ proton and an
  incoming neutrino energy of 1 GeV.  In the left panel, the full
  (dashed) line corresponds to the vector form-factor parameterization of
  Eqs.~(\ref{free11-1}) and (\ref{free11-2}) (Eq.~(\ref{free12})).
  The right panel studies the sensitivity of the cross sections to the
  various parameterizations for $C_5^A(Q^2)$, contained in Fig.~\ref{axial}. }
\label{FFs}
\end{figure}
Before discussing the nuclear effects described in section
\ref{bound}, we address some topics related to the elementary $\Delta$
couplings introduced in section \ref{me-free}.  Figure \ref{FFs}
appraises the sensitivity of the $Q^2$ distribution to uncertainties
residing in the vector and axial-vector form factors.  The left-hand
panel contrasts the widely used M1-dominance parameterization for the
vector form factors with a more recent fit to electroproduction
helicity amplitudes, which extends beyond magnetic-dipole dominance
\cite{Olga2}.  With the Lalakulich fit of Eq.~(\ref{free12}) one
finds cross sections which are about $10\%$ higher than those obtained
with the M1-dominance form factors of Eqs.~(\ref{free11-1}) and (\ref{free11-2}).
The discrepancy between both parameterizations is significant, as
vector form factors are usually regarded as well-known when they are
used as input to extract the far less known axial form factors from
neutrino-scattering data.  As pointed out in section \ref{me-free},
the current situation for the axial-vector form factors is somewhat
more dramatic.  To see how uncertainties in $C_5^A(Q^2)$ affect the
cross section, we have performed computations with both the
phenomenological result of Eq.~(\ref{free13}) and the theoretical
calculations shown in Fig.~\ref{axial}.  The right-hand panel of
Fig.~\ref{FFs} shows what this implies for the $Q^2$ distribution.
Clearly, the $Q^2$ evolution of the $\Delta$-production cross section
exhibits a strong sensitivity to the adopted $C_5^A(Q^2)$
parameterization.  Near $Q^2 = 0$, cross sections using the $\chi$CQ-
and QCD-model results are about $40\%$ lower than the calculation with
the $\Delta$-dominance fit.  This is almost entirely due to the difference in
$C_5^A(0)$ values, which yields a ratio of $(0.9)^2/(1.2)^2 \approx
0.56$ for the dominant cross-section contribution.  The rapid fall-off
predicted by the $\chi$CQ model results in cross-section values that
are much lower over the whole $Q^2$ range.  On the other hand, the QCD
calculation foretells a less steep dipole dependence, leading to more
strength towards higher $Q^2$ values.  The $\chi$CQ result for $C_5^A$
halves the integrated cross section with respect to the calculation
with the $\Delta$-dominance fit.                 
\begin{figure}[t]
\includegraphics[width=9cm]{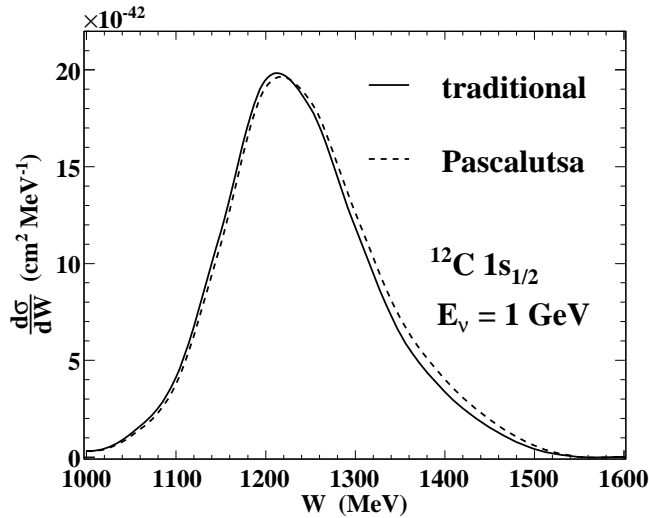}
\caption{ Invariant-mass dependence of the $\Delta^{++}$-production
  cross sections for a $1s_{1/2}\ ^{12}\mbox{C}$ proton and an
  incoming neutrino energy of 1 GeV.  The hadronic invariant mass is
  defined as $W = \sqrt{(k_{\pi}+k_{N})^2}$.  The full (dashed) line uses the
  $\Delta\pi N$ coupling of Eq.~(\ref{free15}) (Eq.~(\ref{free18})). }
\label{Pasc}
\end{figure}
To investigate the impact of different $\Delta$-decay couplings, we
have computed $W$-distributions using both the traditional coupling of
Eq.~(\ref{free15}) and the Pascalutsa coupling of Eq.~(\ref{free18}).
The results are shown in Fig.~\ref{Pasc}, where it can be seen that
differences between the two approaches are small.  Although the
Pascalutsa coupling yields higher values in the tail of the
$W$-distribution, we infer an overall effect that does not exceed the
$2\%$ level.  

\subsection{Nuclear-model effects}
\label{results-pt2}
  
\begin{figure}[t]
\includegraphics[width=18cm]{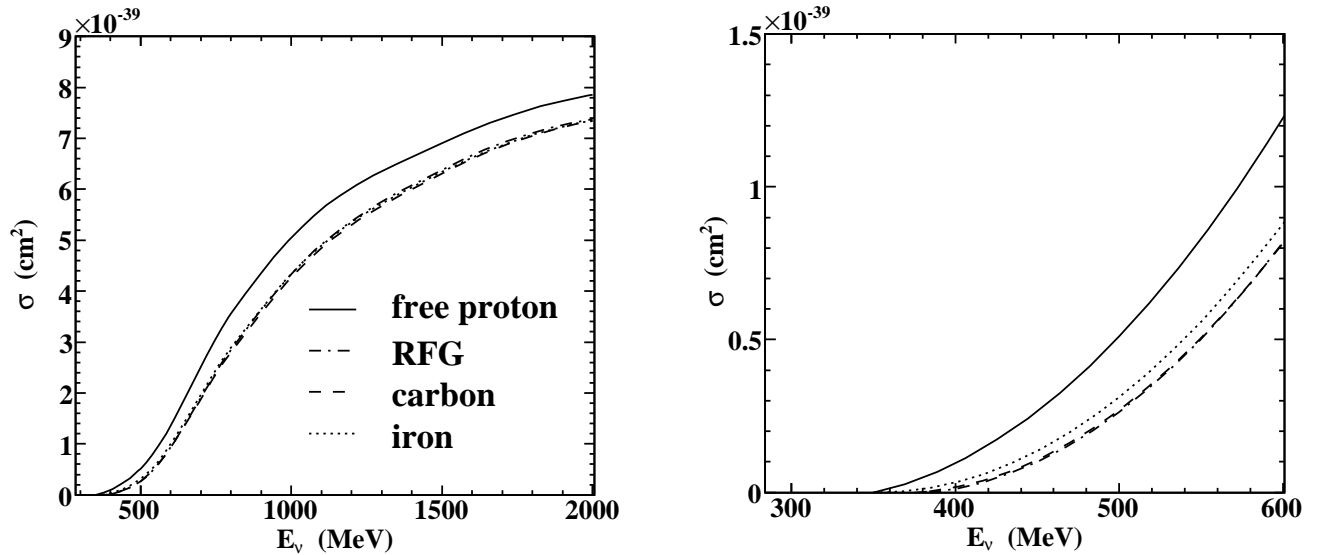}
\caption{ Total cross sections per nucleon for $\nu_{\mu} + p \stackrel{ \Delta^{++} }{ \rightarrow } \mu^{-} + p +
\pi^{+}$.  The full line represents the elementary process, for
scattering from a free proton.  The dash-dotted line stands for the
RFG calculations, whereas the dashed (dotted) line corresponds to
scattering from a carbon (iron) target nucleus.  The right panel
focusses on the threshold region. }
\label{Total}
\end{figure}

\begin{figure}[t]
\includegraphics[width=9cm]{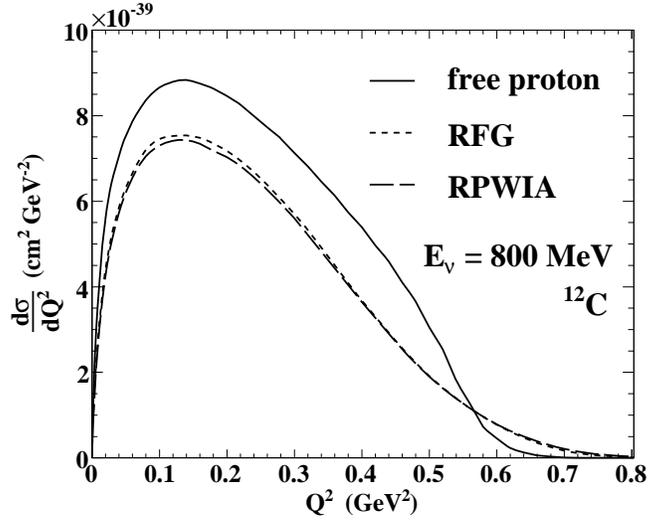}
\caption{ Cross section per nucleon for $\nu_{\mu} + p \stackrel{ \Delta^{++} }{ \rightarrow } \mu^{-} + p +
\pi^{+}$ on carbon at an incoming neutrino energy of $800$\ MeV.  The
full line represents the elementary process, whereas the short-dashed
(long-dashed) line stands for the RFG (RPWIA) calculation.}
\label{Qsq}
\end{figure}

\begin{figure}[t]
\includegraphics[width=9cm]{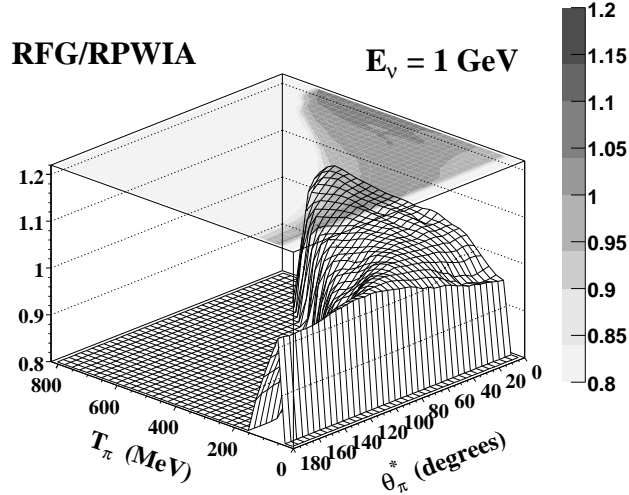}
\caption{ Ratio of RFG to RPWIA computations for the
  $d^2\sigma/dT_{\pi}d\cos\theta^*_{\pi}$ cross section of the process
  $\nu_{\mu} + p \stackrel{ \Delta^{++} }{ \rightarrow } \mu^{-} + p +
\pi^{+}$.  A carbon target and an incoming neutrino energy of $1$\ GeV
are considered. }
\label{Ratio}
\end{figure}

\begin{figure}[t]
\includegraphics[width=9cm]{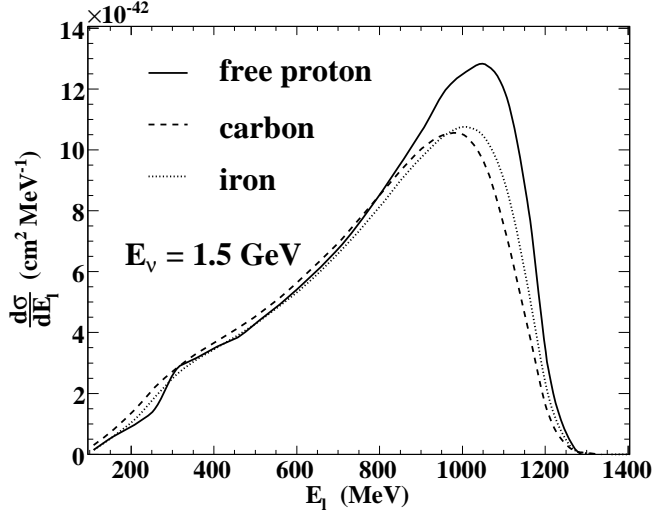}
\caption{ The $\nu_{\mu} + p \stackrel{ \Delta^{++} }{ \rightarrow } \mu^{-} + p +
\pi^{+}$ cross section per nucleon as a function of the lepton energy
$E_l$ for $E_{\nu}=1.5$\ GeV.  The
full line represents the elementary process, whereas the dashed
(dotted) line refers to scattering from carbon (iron).}
\label{CvsFe}
\end{figure}

In this subsection, the results of section \ref{results-pt1} will be
put in a more general perspective.  To this end, we will compare
neutrino-nucleus with neutrino-nucleon cross sections.  Figure~\ref{Total} shows how
the total strength for the process in Eq.~(\ref{res1}) varies with the
incoming neutrino energy.  
Under the same kinematical conditions and with similar input for the
$\Delta$ couplings, our results for the elementary process compare very well with the
predictions published in Ref.~\cite{Hernandez}.                        
Turning to the predictions for target nuclei, Fig.~\ref{Total} shows how the
elementary cross section is halved near threshold.  For higher
incoming energies, the effect dwindles to $20\%$ at $E_{\nu} = 800\
\mbox{MeV}$ and $8\%$ at $E_{\nu} = 2\ \mbox{GeV}$.  Most strikingly, the
RFG calculations are in good to excellent agreement with both the
carbon and iron RPWIA results.  The only discernable feature of
Fig.~\ref{Total} is that the iron curve exceeds the carbon and RFG
ones by roughly $15\%$ just beyond threshold.  This can be
understood after recognizing that the iron result is largely due to
outer-shell protons, which are less bound than the corresponding carbon ones.
Clearly, the nuclear-target cross sections are very sensitive to
binding-energy differences at lower incoming energies.  These effects,
however, vanish at higher neutrino energies and fall to a $1\%$
correction level at $E_{\nu} = 1\ \mbox{GeV}$.  As a matter of fact,
at sufficiently high energies RFG calculations with a well-chosen
binding-energy correction are almost indiscernible from the
corresponding RPWIA results.  These findings
are more detailedly assessed in Figs.~\ref{Qsq}, \ref{Ratio} and \ref{CvsFe}.                
Figures~\ref{Qsq} and \ref{Ratio} compare RFG and RPWIA computations.
The former considers scattering from a carbon target at $E_{\nu} =
800$\ MeV, which corresponds to the mean energy of the neutrino beam
used by the MiniBooNE experiment.  As can be appreciated from
Fig.~\ref{Qsq}, the RFG and RPWIA models produce almost identical
results.  In Fig.~\ref{Ratio}, we present the ratio of RFG to carbon
RPWIA results for the twofold cross section
$d^2\sigma/dT_{\pi}d\cos\theta^*_{\pi}$, where $T_{\pi}$ is the
outgoing pion's kinetic energy and $\theta^*_{\pi}$ its scattering
angle relative to the neutrino-beam direction.  Apart from the
threshold region, where numerical instabilities induce large
fluctuations, it is observed that differences between the RFG and
RPWIA result do not exceed the $5\%$ level over the whole
$(T_{\pi},\theta^*_{\pi})$ range.  In addition, the largest deviations
occur where the cross section has hardly any strength.  Consequently,
upon integrating over $T_{\pi}$ and $\theta^*_{\pi}$, we find that the total
RFG cross section exceeds the RPWIA one by about $2\%$.          
Figure~\ref{CvsFe} compares the cross section for a carbon nucleus
with the one for an iron nucleus at $E_{\nu} = 1.5$\ GeV.  Although the
total strength, integrated over the outgoing muon energy $E_l$, is the
same for both nuclei, it is interesting to note that the iron
distribution is shifted with respect to the carbon cross section.
Again, this reflects the fact that a carbon proton requires, on
average, more energy than an iron proton to be knocked out of the
nucleus, leaving therefore less energy for the outgoing muon.  
\begin{figure}[t]
\includegraphics[width=9cm]{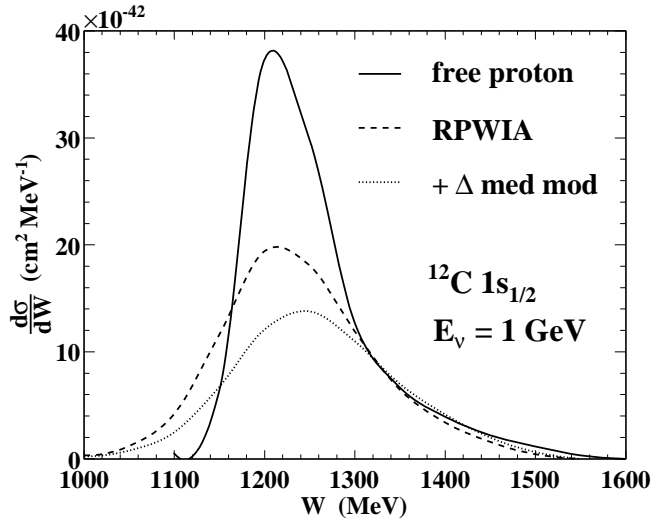}
\caption{Invariant-mass distribution for $\nu_{\mu} + p \stackrel{ \Delta^{++} }{ \rightarrow } \mu^{-} + p +
\pi^{+}$ on a $1s_{1/2}\ ^{12}\mbox{C}$ proton at an incoming neutrino
energy of $1$\ GeV.  The full (dashed) line represents the calculation
for a free (bound) proton.  The dotted curve adds the effect of
$\Delta$ medium modifications to the RPWIA result.}
\label{Delta_MM}
\end{figure}
To assess the influence of medium modifications to the mass and width of
the $\Delta$, we have computed the $W$-distribution according to the
prescription given in Eq.~(\ref{MM1}).  In Fig.~\ref{Delta_MM}, we
contrast the elementary cross section with RPWIA calculations for
carbon.  Compared to the free case, the RPWIA
result that does not include $\Delta$ medium modifications is seen to
be heavily suppressed at the $\Delta$ pole.  The nuclear binding
brings about a broadening of the $W$-distribution, mainly reallocating
strength to lower invariant masses.  When medium modifications are
taken into account, the $\Delta$ pole is shifted towards higher $W$
values, by an amount that corresponds to the mass shift in
Eq.~(\ref{MM1}).  Compared to the plain RPWIA case, the peak is again
suppressed and broadened, owing to the increased medium-modified
width.  On the whole, we observe a $25\%$ reduction of the RPWIA cross
section when $\Delta$ medium modifications are included.                        

\subsection{Results under MiniBooNE and K2K kinematics}

\begin{figure}[t]
\includegraphics[width=9cm]{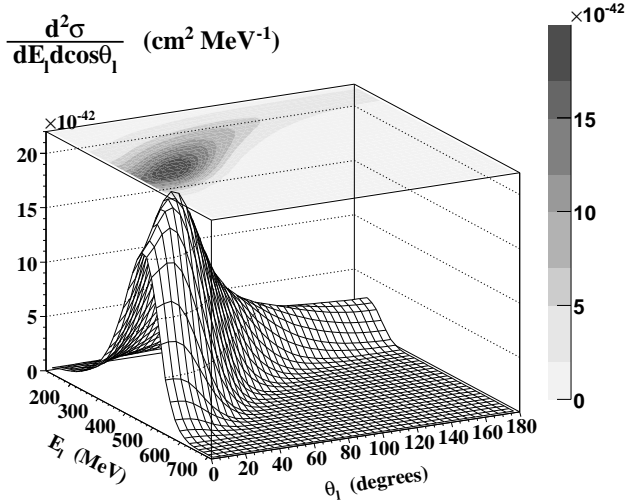}
\caption{ Cross section per nucleon for $\nu_{\mu} + p \stackrel{ \Delta^{++} }{ \rightarrow } \mu^{-} + p +
\pi^{+}$ against outgoing-muon energy and scattering angle.  The
incoming neutrino energy is $800$ MeV, the target nucleus is carbon. }
\label{MiniBooNE-twofold}
\end{figure}

\begin{figure}[t]
\includegraphics[width=18cm]{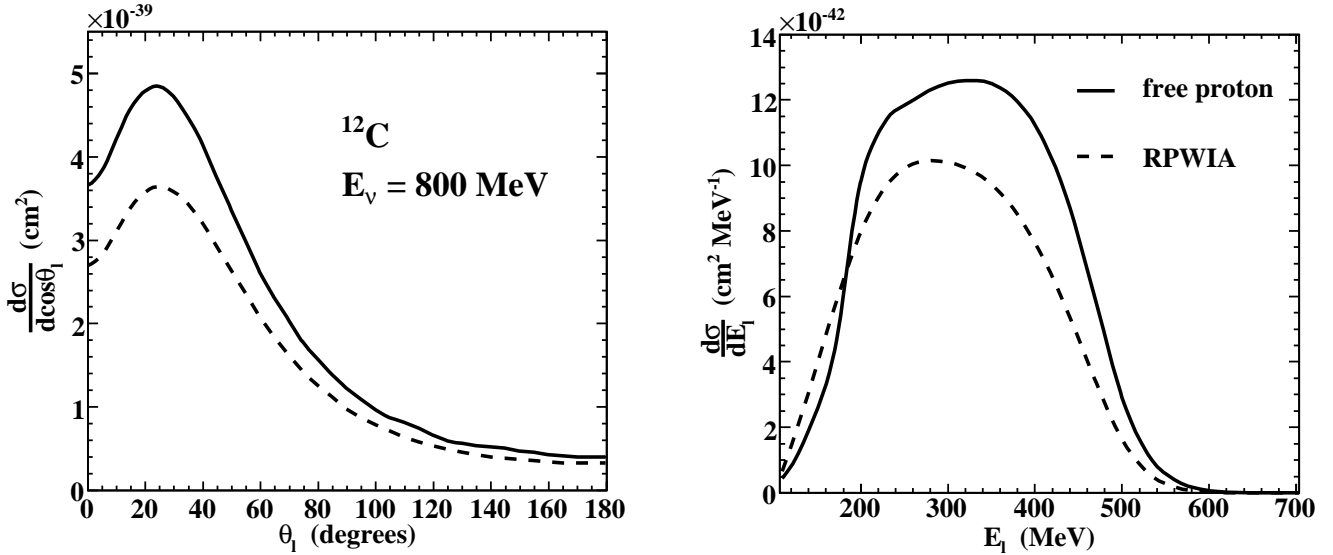}
\caption{ Cross sections per nucleon for $\nu_{\mu} + p \stackrel{ \Delta^{++} }{ \rightarrow } \mu^{-} + p +
\pi^{+}$, for $800$\ MeV neutrinos scattering from a carbon target.
The left (right) panel shows the cross section as a function
of the outgoing-muon scattering angle (energy).  Each of the panels
contrasts the elementary cross section (full line) with the RPWIA
result, without $\Delta$ medium modifications (dashed line). }
\label{MiniBooNE-onefolds}
\end{figure}

\begin{figure}[t]
\includegraphics[width=9cm]{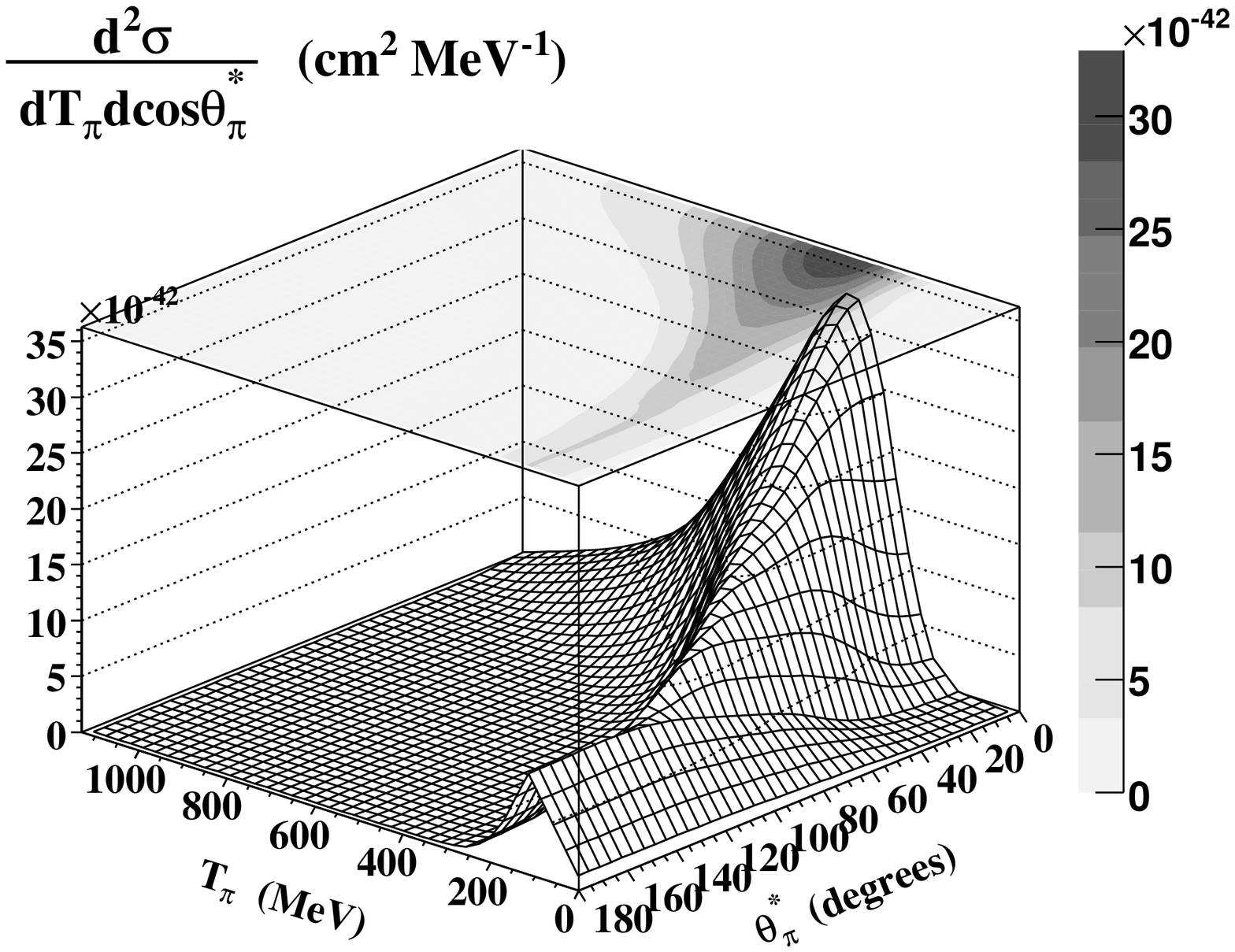}
\caption{ Cross section per nucleon for $\nu_{\mu} + p \stackrel{ \Delta^{++} }{ \rightarrow } \mu^{-} + p +
\pi^{+}$ against outgoing-pion kinetic energy and scattering angle.  The
incoming neutrino energy is $1.3$ GeV, the target nucleus is oxygen.}
\label{K2K-twofold}
\end{figure}

\begin{figure}[t]
\includegraphics[width=18cm]{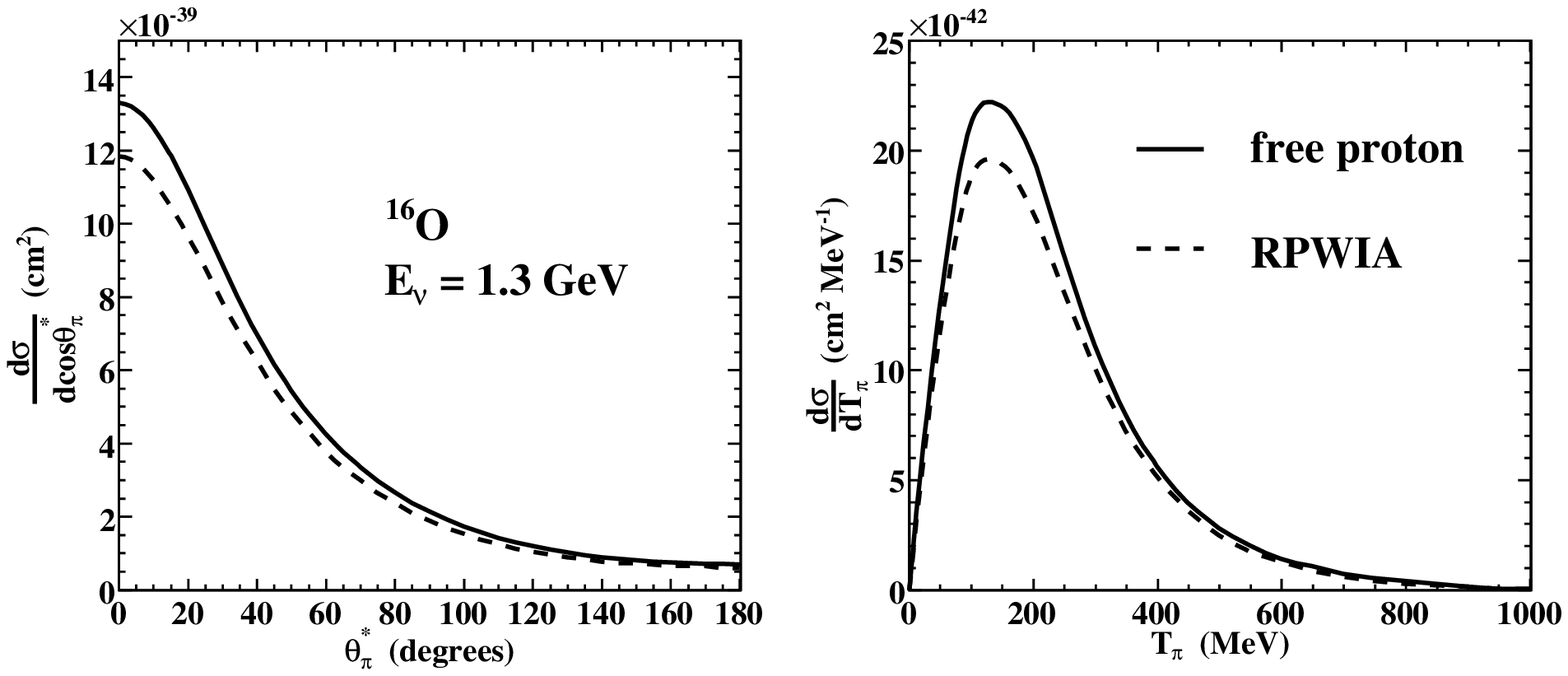}
\caption{ Cross sections per nucleon for $\nu_{\mu} + p \stackrel{ \Delta^{++} }{ \rightarrow } \mu^{-} + p +
\pi^{+}$, for $1.3$\ GeV neutrinos scattering from an oxygen target.
The left (right) panel shows the cross section as a function
of the outgoing-pion scattering angle (kinetic energy).  Each of the panels
contrasts the elementary cross section (full line) with the RPWIA
result, without $\Delta$ medium modifications (dashed line).}
\label{K2K-onefolds}
\end{figure}

In view of recent results presented by the MiniBooNE and K2K
collaborations \cite{NuInt1}, we conclude this section with some
computations for the specific neutrino energies and target nuclei
employed by these experiments.  From an experimental viewpoint, the
most accessible distributions are the ones with respect to
outgoing-muon variables.  Fig.~\ref{MiniBooNE-twofold} depicts an
RPWIA calculation for a two-fold differential cross section against
the outgoing-muon energy and scattering angle with respect to the
neutrino beam.  The incoming neutrino energy is fixed at $800$ MeV,
corresponding to MiniBooNE's mean beam energy.  Since MiniBooNE has
carbon as target material, this calculation was performed on a
carbon nucleus.     
The result shown in Fig.~\ref{MiniBooNE-twofold} can be integrated
over $\theta_l$ or $E_l$ to yield the one-fold cross sections
displayed in Fig.~\ref{MiniBooNE-onefolds}.  Relative to the free cross section,
the angular distribution for a carbon target is evenly reduced by
about $20\%$.  In general, the outgoing muon prefers a forward
direction, although a minor shift seems to take place between the free
and the bound case.  This effect relates to the change in the muon-energy
distribution, depicted in the right-hand panel of
Fig.~\ref{MiniBooNE-onefolds}.  Indeed, for scattering off bound
protons, one observes a shift of the $E_l$ distribution towards lower
values.  Recognizing the correlation between high muon energies
and forward scattering angles, as can be appreciated in
Fig.~\ref{MiniBooNE-twofold}, the bound case will correspondingly
yield a larger number of events at slightly higher scattering angles.
We also note that the RPWIA result fades out sooner than the
elementary cross section, because a certain amount of energy is needed
to knock the carbon proton out of its shell.      
Planned experiments like MINER$\nu$A endeavor to have a good energy
resolution for both the muon and the hadronic final state.  The
ability to detect the outgoing pion or nucleon or even both would
allow a detailed study of different nuclear effects.  In
Figs.~\ref{K2K-twofold} and \ref{K2K-onefolds} we present cross sections
versus the pion kinetic energy $T_{\pi}$ and pion scattering angle
relative to the beam direction $\theta^{*}_{\pi}$.  This time, we adopted K2K settings,
namely an oxygen target hit by neutrinos with an energy of $1.3$\
GeV.           
From the left-hand panel of Fig.~\ref{K2K-onefolds}, one infers that,
within the RPWIA model, the outgoing pion prefereably leaves the
nucleus along the beam direction.  As for the kinetic-energy
distribution, we observe a comparable reduction and shift of the
strength as in the muon-energy distribution.

\section{ Conclusion and outlook }
\label{conclusion}
 
We have developed a relativistic framework to study $\Delta$-mediated
one-pion production from nuclei at medium energies.  The proposed
formalism offers great flexibility in calculating various observables 
both for the free process and for scattering from nuclear targets.
Motivated by operational and planned experiments, we have conducted a 
systematic study by addressing the impact of $\Delta$-coupling
ambiguities on the $Q^2$ and $W$ distributions.  Cross sections are
found to vary by as much as $10\%$ depending on whether or not the
M1-dominance assumption is used to extract the vector form factors.
This is very significant, as the extracted value for the axial mass
$M_A$ depends heavily on the model applied in the analysis of the
neutrino-scattering data and, therefore, on a reliable input for the
vector form factors.  Uncertainties in the dominant axial form factor,
$C_5^A(Q^2)$, have a dramatic effect on the $\Delta$-production cross
sections.  At low $Q^2$, a $25\%$ reduction of the off-diagonal
Goldberger-Treiman value $C_5^A(0) = 1.2$ leads to cross sections that are smaller by
$40\%$.  In the case of a $Q^2$ dependence that is steeper than a
modified-dipole form, the effect increases to almost $50\%$
over the whole $Q^2$ range.  
In the $W$ distribution, we observe $2\%$-level deviations between the
traditional $\Delta$-decay coupling choice and a consistent one, which
effects a decoupling from the spin-1/2 terms.  To investigate the influence of nuclear effects, we have
computed RPWIA neutrino-nucleus cross sections for carbon, oxygen and
iron nuclei.  We have briefly touched on the topic of $\Delta$
medium modifications.  Using a prescription that gives good results in
photo-induced two-nucleon knockout and electron-scattering studies, we
infer a $20$-$25\%$ suppression of the RPWIA cross sections due to medium effects.  The nuclear responses
are very sensitive to binding-energy differences at lower neutrino
energies.  From $E_{\nu} = 1$\ GeV onwards, the cross sections per
nucleon for different nuclear targets are seen to agree at the $1\%$
level.  To assess the nuclear-model uncertainty in our description of 
$\Delta$-mediated one-pion production, we have also contrasted the
RPWIA results with calculations performed within an RFG model with a
well-considered binding-energy correction.  At $1$-GeV neutrino
energies, differences between one- and two-fold distributions computed
within both models do not exceed the $5\%$ level.  The agreement is
better for total cross sections, where deviations between the RFG and
RPWIA model dwindle to $1$-$2\%$.  Hence, for sufficiently high
incoming neutrino energies, the influence of Fermi motion, nuclear
binding and the Paul exclusion principle can be well described by
adopting an RFG model with binding-energy correction.  The RFG
model, however, just as the RPWIA approach, falls short in implementing FSI and
nuclear correlations of the short and long-range type.  Contrary to the
RFG, the model proposed in this work has the important advantage that
it can serve as a starting point for a relativistic and
quantum-mechanical study of FSI mechanisms.  As a matter of fact, the
inclusion of FSI for the ejected pions and nucleons is currently under
study.  To this end, we closely follow the lines of
Ref.~\cite{Cosyn1}, where use is made of a relativistic Glauber model for fast
ejectiles and an optical-potential approach for lower ejectile energies.

\acknowledgments
The authors acknowledge financial support from the Research Foundation
- Flanders (FWO), and the Research Council of Ghent University.

\end{document}